# High-pressure behavior of methylammonium lead iodide (MAPbI$_3$) hybrid perovskite

*Francesco Capitani,[a] Carlo Marini,[b] Simone Caramazza,[c] Paolo Postorino,[c] Gaston Garbarino,[d] Michael Hanfland,[d] Ambra Pisanu,[e] Paolo Quadrelli,[e] Lorenzo Malavasi [e,*]*

[a]Synchrotron SOLEIL, L'Orme des Merisiers, Saint-Aubin, 91192 Gif-sur-Yvette, France; [b]CELLS-ALBA, Carretera B.P. 1413, Cerdanyola del Valles 08290, Spain; [c]Department of Physics, University "Sapienza", Rome, Italy; [d]European Synchrotron Radiation Facility, Grenoble Cedex, France; [e]Department of Chemistry and INSTM, University of Pavia, Pavia, Italy;

**Corresponding Author**

*Lorenzo Malavasi, Department of Chemistry, University of Pavia, Pavia, Italy,

lorenzo.malavasi@unipv.it

**ABSTRACT**

In this paper we provide an accurate high-pressure structural and optical study of MAPbI$_3$ hybrid perovskite. Structural data show the presence of a phase transition towards an orthorhombic structure around 0.3 GPa followed by full amorphization of the system above 3 GPa. After releasing pressure the systems keeps the high-pressure orthorhombic phase. The occurrence of these structural transitions is further confirmed by pressure induced variations of the photoluminescence signal at high pressure. These variations clearly indicate that the bandgap value and the electronic structure of MAPI change across the phase transition.

# INTRODUCTION

Organometal halide perovskites based solar cells represent one of the most exciting and promising technology in the field of photovoltaics, with certified efficiency that reached about 20% in few years.[1-4] Most of the investigated cells are based on the methylammonium (MA) lead iodide (MAPbI$_3$) absorbing material which adopts a tetragonal perovskite structure with space group *I*4/*mcm* at room temperature,[5] with the methylammonium ion sitting on the perovskite A-site and the Pb on the B-site. MAPbI$_3$ presents different phase transitions as a function of temperature: i) above 327.4 K it adopts a cubic structure (*Pm-3m*); ii) between 162.2 K and 327.4 K it is tetragonal (*I*4/*mcm*) and below 162.2 K it is orthorhombic (*Pna*2$_1$).[5] The crystal structure of MAPbI$_3$ has been thoroughly investigated as a function of temperature with various techniques. In particular, a recent neutron diffraction study on a fully deuterated sample, could determine the complete structure and cation orientation in methylammonium lead iodide in the 100-352 K range, providing a clear picture of the increasing positional disorder by increasing temperature.[6]

On the other hand, the behavior of lead halide perovskites under applied pressure has been less extensively investigated. The response to pressure of perovskites is a very important topic of interest for the solid-state chemistry and physic, being an intriguing tool to tune the structural and physical properties of this class of materials. Swainson *et al.* reported a high-pressure (HP) neutron diffraction study on deuterated MAPbBr$_3$ showing a transformation from *Pm-3m* to *Im*-3 symmetry occurring at about 1 GPa, followed by amorphization around 2.8 GPa without the cations undergoing long-range orientational ordering.[7] In the same work, the authors applied pressure at room temperature and cooled down the perovskite entering the low-temperature orthorhombic phase (*Pnma*) showing the persistence of the *Im*-3 phase even at 100 K.[7] A very recent x-ray diffraction HP study on the same material, reported two phase transformations below 2 GPa (from *Pm*-3*m* to *Im*-3, then to *Pnma*) and a reversible amorphization starting from about 2 GPa, that was attributed to the tilting of the PbBr$_6$ octahedra and destroying of long-range ordering of MA cations, respectively.[8] In the same work, the authors observed an anomalous evolution of the band-gap during compression with red-shift followed by blue-shift. No pressure transmitting medium was used in this work.

Very recently, MAPbI$_3$ nanorods have been object of high-pressure *in situ* photocurrent, impedance spectroscopy, and x-ray diffraction (XRD) measurements by Ou et al.[9] Main result of this work was the evidence of an anomalous change in the electrical transport parameters, relaxation frequency, and relative permittivity of MAPbI$_3$ nanorods around 0.6 GPa due to the occurrence of a tetragonal to orthorhombic phase transition.[9] Amorphization of the sample is observed above ca. 20 GPa.[9] However, in this work the authors did not employed any pressure transmitting medium and the XRD analysis was afforded in a qualitative way without reporting any lattice parameters value and trend as a function of the applied pressure. A more comprehensive study on the HP structural and optical properties of MAPbI$_3$ was reported in arxiv by Wang *et al*.[10] Main result of this work was the observation of a phase transition from tetragonal to orthorhombic phase around 0.3 GPa followed by partial amorphization of the sample up to the maximum *P*-range explored (ca. 7 GPa) and recovery of the tetragonal phase after releasing pressure. In ref. 10 amorphization of the sample is claimed but clear and intense diffraction peaks are visible up to the highest pressure considered. Moreover, no information about the lattice parameters trend by reducing pressure are given in this pre-print and so no information about a possible hysteresis are provided. Moreover, MAPbI$_3$ employed in this study was a commercial sample and no information on its basic characteristics and quality are available in the paper.

From the above discussion it is clear that there is a significant spread in the reported properties of MAPbI$_3$ behavior at high-pressure and a possible close dependence with samples employed in the studies. In addition, a partial lack of experimental data (*e.g.*, structural results by reducing pressure) and/or doubts on the obtained results (*e.g.*, measurements performed without pressure transmitting medium) suggest that a rigorous and accurate investigation on MAPbI$_3$ as a function of pressure is still needed.

For this reasons, in the present work we report a detailed HP investigation on a well characterized sample of MAPbI$_3$ perovskite carried out by means of HP x-ray diffraction (XRD) and photoluminescence (PL) employing diamond anvil cells (DAC).

**EXPERIMENTAL SECTION**

MAPbI$_3$ has been synthesized starting from a proper amount of Pb acetate (99.9%, Aldrich), which was dissolved in an excess of HI acid (55% in water, Aldrich) under nitrogen atmosphere and stirring. The solution was heated to 100°C and the methylammonium (40 wt% in water, Aldrich) is added in equimolar amount. A precipitate was formed immediately after the amine addition. The solution is then cooled down to 46°C at 1°C/min, and the precipitate is immediately filtered and dried at 60°C under vacuum overnight.

High-pressure X-ray diffraction data were collected on the ID09 beamline at the ESRF Facility in Grenoble with a diamond anvil cell (DAC). The beam ($\lambda$ = 0.414101 Å) size on the sample is normally about 10 × 10 μm$^2$. A DAC was employed with low-fluorescence IA, 600-μm culet diamonds. The samples were finely milled and a high pressure DAC loading, with He as pressure transmitting medium, was performed. The image plate detector is a Mar555 reader.

Photoluminescence (PL) spectra were collected using a Horiba LabRAM HR Evolution microspectrometer in backscattering geometry. Samples were excited by the 632.8 nm ($\upsilon_0$ = 15803 cm$^{-1}$) radiation of a He-Ne laser with 30 mW output power. A long-working-distance 20X objective has been used to collect the PL signal from the sample in the DAC. Under this condition a ~ 3 μm laser spot at the sample surface allowed for a careful homogeneity check of both pressure and sample. The diamond anvil cell was equipped with IIA diamonds (400 μm culet diameter) and the sample loaded in a hole 200 μm diameter and 50 μm height drilled in a steel gasket. NaCl was used as a pressure transmitting medium as in the case of other pseudocubic perovskite compounds (see, *e.g.*, ref. 11) and the ruby fluorescence technique was exploited to measure the pressure in situ.

**RESULTS AND DISCUSSION**

MAPbI3 sample used in the following experiments has been fully characterized to check its phase purity and properties. Figure S1 in the Supporting Information[12] shows the differential thermal analysis (DSC) graph of MAPbI3 indicating the tetragonal to cubic phase transition at about 54°C as reported in the current literature.[5] Morphology of the sample has been checked by means of scanning electron microscopy (SEM) and is shown in Figure S2.[12] The samples is made of micron-sized particles of spheroidal shape. Before the HP-XRD experiment, the samples have been carefully grinded in order to optimize the averaging of diffraction.

Figure 1 shows the refinement of the XRD pattern collected at *P*=0 GPa (open DAC cell) on the methylammonium lead iodide sample used in the experiments reported in this work.

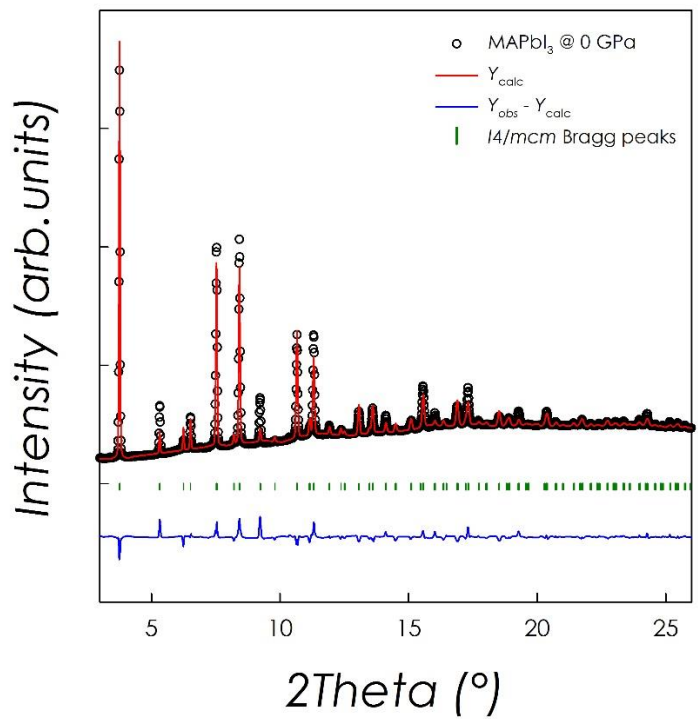

**Figure 1:** Fit of the MAPbI3 perovskite at ambient pressure.

As can be appreciated from Figure 1, the sample is single phase and refinable in the *I4/mcm* space group. Lattice parameters obtained from the fit are *a*=*b*=8.909(1) Å and *c*=12.620(3) Å, in agreement with literature data.[5]

After confirming the quality of the sample, XRD measurements have been carried out by increasing pressure in a DAC by using He as pressure transmitting medium. Figure 2 reports selected HP XRD patterns collected during the pressure increase from ambient pressure (0 GPa) to 3.65 GPa every ca. 0.3 GPa, after pressure stabilization for about 60 minutes. Such narrow pressure increase steps and equilibration allows to obtain highly reliable data and are required when studying highly compressible systems such as hybrid perovskites.[13] For example, in ref. 10 bigger *P*-increase steps of about 0.5-1 GPa has been used in the XRD experiments.

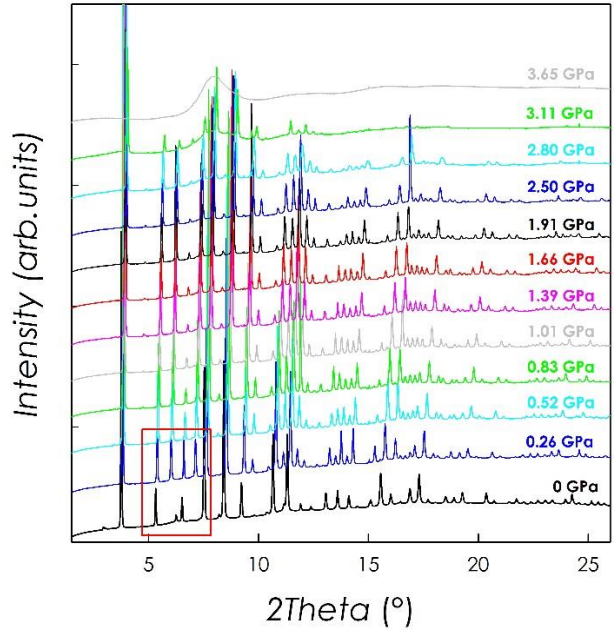

**Figure 2:** HP XRD patterns for the MAPbI$_3$ perovskite. Red box highlights a representative region where the phase transition occurs.

From Figure 2 it is possible to observe a significant change in the XRD pattern from the ambient pressure structure (tetragonal, s.g. *I*4/*mcm*, *a*=*b*=8.909(1) Å, *c*=12.620(3) Å) already at 0.26 GPa. With the further increase of pressure, no other significant peaks and/or peaks splitting are observed in the patterns. At about 2.80 GPa, a relevant contribution from the background is visible which is a sign of the formation of an amorphous phase which is then the main phase at 3.65 GPa, thus indicating the complete amorphization of the systems at that pressure. At his pressure, essentially no peaks are detected in the XRD pattern (Figure S3).[12] A better indication of the range of amorphous transition has been obtained from the analysis of the structural data (see later in the text).

The result reported in our work concerning the set-up of an amorphous phase for MAPbI$_3$ at about 3.65 GPa is significantly different with respect to the data reported in ref. 9 where full amorphization of MAPbI$_3$ is achieved at about 21 GPa. Such big difference may be related to the peculiar nanostructure of the sample used in ref. 9 (nanorods) but, most probably, could be correlated to the absence of any pressure transmitting medium which casts significant doubts on the results reported particularly in the case of cylindrical shaped nanorods. On the other hand, in ref. 10, clear diffraction peaks are still detected at the maximum pressure employed in this work (7 GPa) even though the authors claim that the system became amorphous. Wang and co-workers reports a nearly full amorphous pattern on the MAPbBr$_3$ at very high pressures above 20 GPa.[8] Also in this work, the authors did not used any *P*-transmitting medium, thus suggesting, as in the case of ref. 9, that these structural data may be strongly dependent upon the absence of hydrostaticity in their measurements.

Concerning the new phase appearing at 0.26 GPa, the structural analysis we carried out shows that it can be nicely refined in the *Imm2* orthorhombic space group as reported in Figure 3.

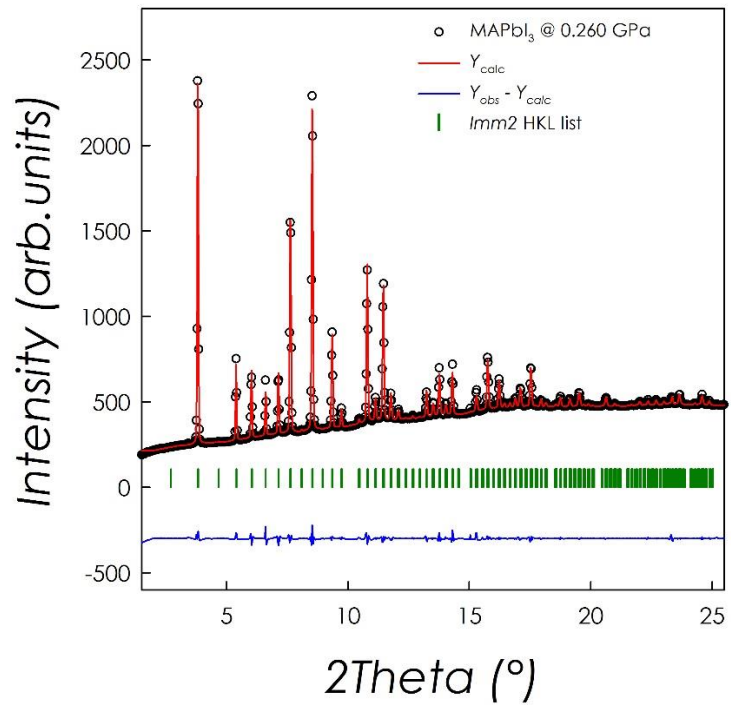

**Figure 3:** Fit of the MAPbI$_3$ perovskite at 0.26 GPa with the *Imm2* space group.

The refined lattice parameters at this pressure are $a$=12.414(1), $a$=12.439(1) and $a$=12.473(1) Å thus indicating that the orthorhombic distortion is relatively small. The low-temperature orthorhombic phases found in lead halide perovskites, namely *Pna*2$_1$ and *Pnma*, were tested but could not account for the observed features in the patterns.

The structure evolution under compression to an orthorhombic phase was previously observed in the MAPbBr$_3$ perovskite by Wang and co-workers.[8] For the methylammonium lead bromide, however, two phases are observed by increasing pressure: the *Im*-3 phase at 0.4 GPa and the *Pnma* phase at 1.8 GPa.[8] For the MAPbI$_3$ both available literature reports indicate the occurrence of the same phase transition we reported in this work at about 0.3 GPa, in agreement with our results (even though in ref. 9 no structural analysis is provided and the symmetry of the new phase is taken from the data of ref. 10).[9,10] It is interesting to observe that, while this low-pressure phase transition occurs at the same pressure for all the experiment carried out so far on the MAPbI$_3$, *i.e.* refs. 9 and 10 and present work, then the behavior at higher pressures is markedly different as discussed previously with reference to the amorphization process.

We also underline that the evolution to an orthorhombic phase by applying pressure is similar to the behavior observed by reducing temperature, even though the unit cell symmetries are different (*Imm*2 by increasing pressure and *Pna*2$_1$ by reducing temperature).

Figure 4 shows the trend of the lattice volume, *V*, as a function of pressure (*a*, *b* and *c* lattice parameters trend is in the Figure S5, together with the list of the values obtained in the refinement, S4)[12] during sample compression (black circles) and decompression (blue circles, see later in the text). We fit data using a 3$^{rd}$ order Birch-Murnaghan equation of state and the obtained results are shown in Figure 3 as red line. In order to get a better constraint value of $V_0$ we used the volume refined after decompression. We remark that no structural data are reported in ref. 9, while in ref. 10 only the compression *V* trend is shown without providing any lattice parameters value which, however, are data of great relevance and usefulness for any theoretical and further experimental investigation on the HP behavior of MAPbI$_3$ perovskite.

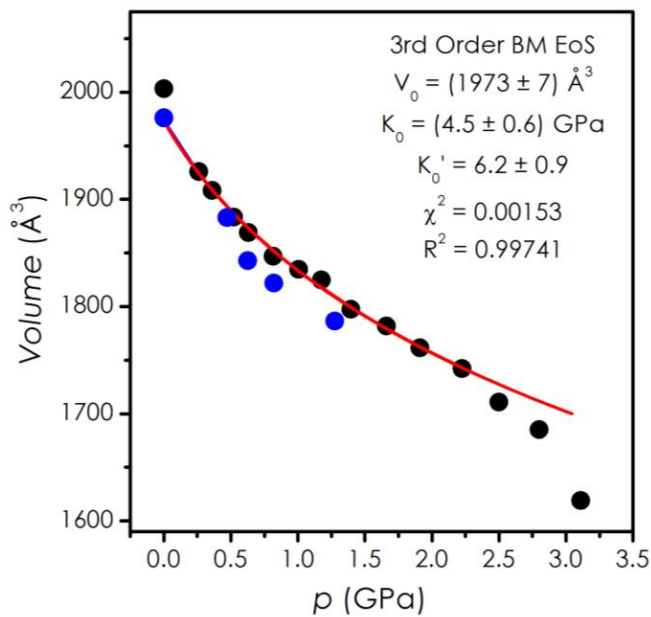

**Figure 4:** Cell volume trend during compression (black circles) and decompression (blue circles). Red line is the Birch-Murnaghan fit of the data.

From the result presented in Figure 4, and in particular from the change in the slope of the data trend and fitting results, it is possible to say that the amorphous transition starts at about 2.3 GPa and is complete at around 3 GPa. The low value of the bulk modulus ($K_0$ = 4.5(6) GPa) for the MAPbI$_3$ perovskite is in agreement with previous data on Sn-based halide perovskites demonstrates the highly compressible nature of these hybrid halide compounds.[13]

In order to check the stability of the amorphous phase, XRD patterns on MAPbI$_3$ were acquired by releasing pressure and Figure 5 shows the evolution of the XRD patters by decreasing pressure from about 3 GPa to ambient pressure.

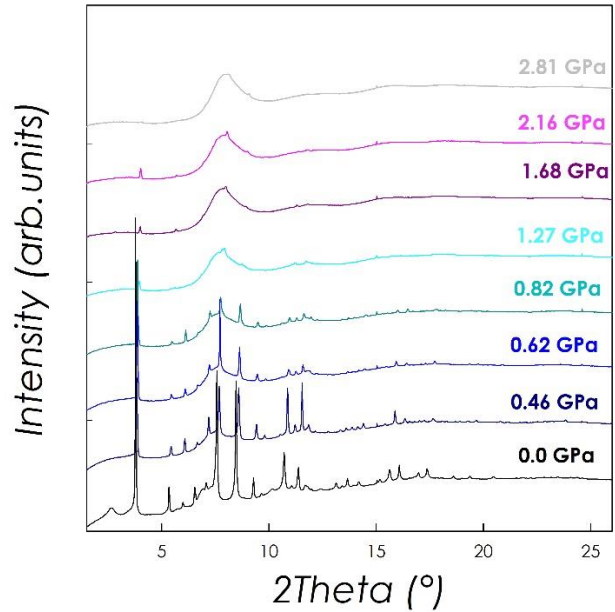

**Figure 5:** HP XRD patterns for the MAPbI$_3$ perovskite by decreasing pressure.

As can be appreciated from Figure 5, the amorphous phase is the main stable phase up to 1.27 GPa while below this pressure, a significant crystalline component appears and grows by further reducing pressure. Even at ambient pressure, the recovery of the crystalline phase is not complete and a contribution of the amorphous phase is still evident, in particular in the region around 8°. Let us remember that, by increasing pressure, the amorphous phase became the main phase above 3 GPa while on the data shown in Figure 5 the amorphous phase is well visible up to low pressures thus indicating a significant metastability of the amorphous phase by reducing *P*. In the previous HP-XRD investigations on both the MAPbBr$_3$ and MAPbI$_3$ perovskites, the authors just reported the diffraction patterns after the full pressure release thus not providing any information about amorphous phase stability during pressure release.[8-10]

The refinement of the ambient pressure XRD after releasing pressure indicates that the stable phase is the high-pressure orthorhombic phase as shown by the XRD refinement of Figure 6.

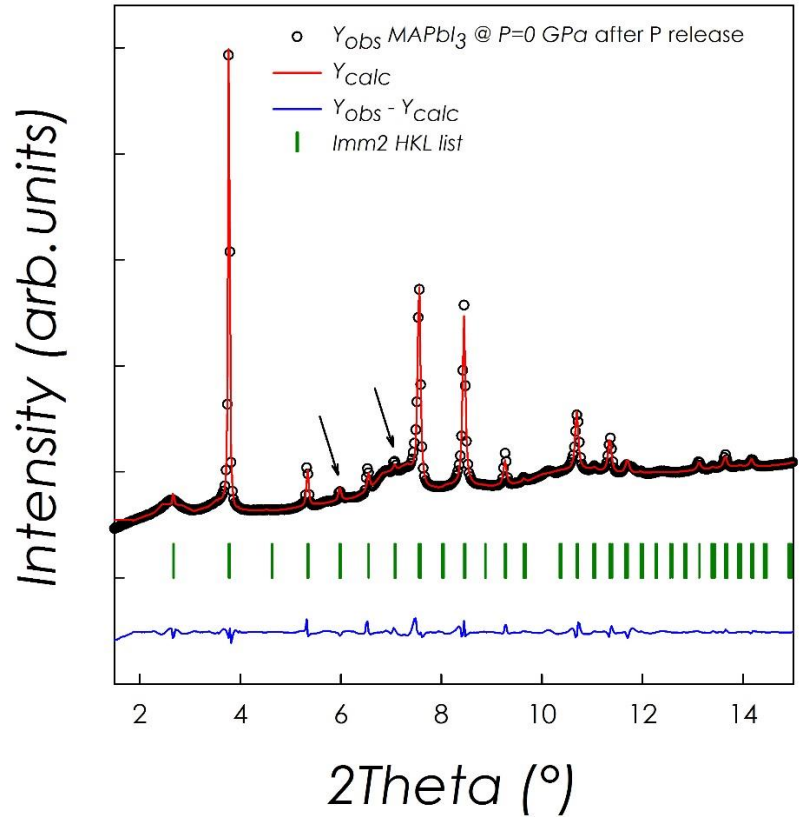

Figure 6: Rietveld refinement of the MAPbI$_3$ perovskite at ambient pressure after releasing pressure with the *Imm2* space group.

In particular, the tetragonal *I4/mcm* phase cannot describe several peaks present in the pattern such as those highlighted – as a selected example - with arrows in Figure 6. The lattice parameter of the recovered phase at 0 GPa are *a*=12.572(1) Å, *b*=12.511(1) Å, and *c*=12.563(1) Å and the cell volume 1976.01(1) Å$^3$. This value is smaller than the cell volume of the as-prepared MAPbI$_3$ of 2003.10(1) Å$^3$ leading to a slightly compressed phase at ambient pressure after pressure release. This result is different with respect to the results found in refs. 9 and 10 for the MAPbI$_3$ where the authors had a full recovery of the crystalline phase with the same symmetry of the starting sample (*I4/mcm*). However, as mentioned before, in ref. 9 no transmission medium was used in the experiment, which may affect the relative stability of the phase(s), while in ref. 10 the authors never achieved a real amorphous phase and the same arguments presented above about the *P*-increase study holds in this case.

Pressure dependent PL spectra have been measured over the pressure range 0 - 5.3 GPa collecting the signal from different points on the sample surface, as enabled by the high instrumental spatial resolution. By this procedure, we could provide a reliable evaluation of the experimental uncertainty, accounting for both weak sample inhomogeneities and small pressure gradients. Spectra collected from a given point at different pressures are shown in Figure 7. In full agreement with previous measurements at *P*=0, the PL signal consists of a single peak with a maximum at 785 ± 5 nm (1.58 ± 0.01 eV).[10] The position of this peak, which is ascribed to near-band-edge transitions,[14-19] matches well the value of the ambient pressure band-gap energy of tetragonal MAPbI$_3$.[14-19] As the pressure is increased, the PL peak position at first shifts remarkably down to 744 nm (1.67 eV) at *P* = 1 GPa, and then slightly moves up to 748 nm (1.66 eV) at *P* = 2.1 GPa. At P=3.3 GPa the main peak moves back to low wavelengths, 736 ± 2 nm (1.685 ± 0.005 eV), but the most striking result is that a new strong peak appears with a maximum at about 690 nm (1.80 eV), which reveals the presence of a new emission mechanism. We notice that these two luminescence features closely resembles those observed at ambient pressure in the orthorhombic phase at temperature below 130 K and extensively discussed in Ref 10.

Upon further increasing the pressure up to *P* = 4.3 GPa, the shape of the PL signal changes completely and a large, weak peak, apparently centered at even lower wavelength, appears. The new shape of the PL signal appears stable under further compression.

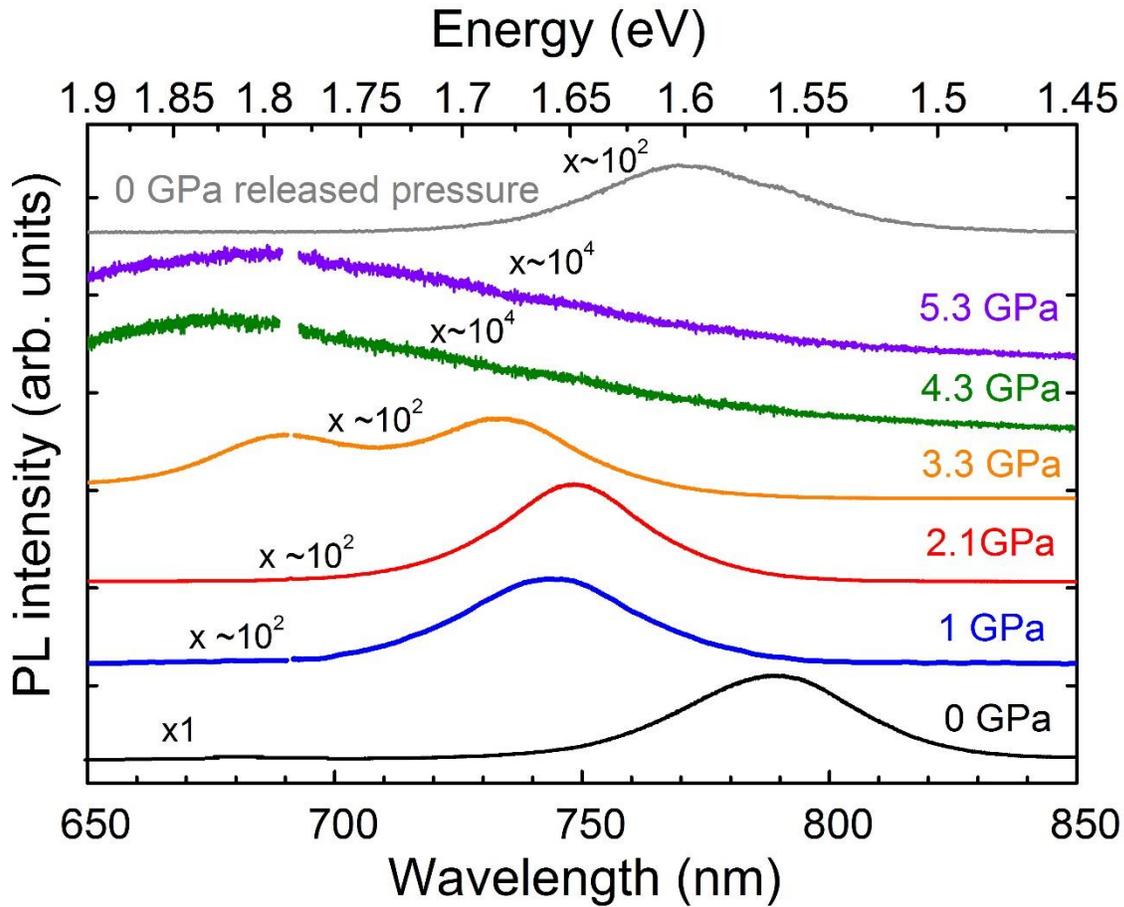

**Figure 7**: Photoluminescence spectra under pressure (*P*= 0 GPa, 1 GPa, 2.1 GPa, 3.3 GPa, 4.3 GPa and 5.3 GPa) collected from a given point. The Raman signal from the diamond anvils at about 690 nm has been removed.

The visual inspection of the sample, provided by optical microscope, basically reflects the pressure induced gap opening shown by the PL signal. Sample microphotographs collected under back illumination at the different pressures are shown in Figure 8. At ambient pressure the sample is actually black but its color progressively drifts toward a reddish appearance as the pressure is increased.

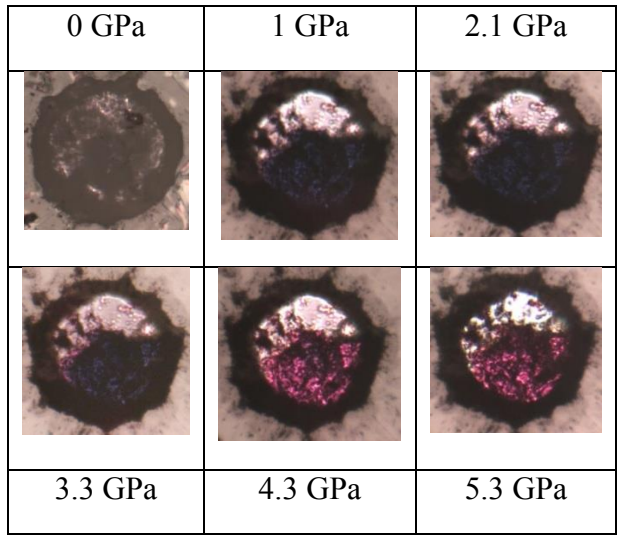

**Figure 8**: Micro-photographs of MAPbI$_3$ inside the DAC at different pressure values.

It is interesting to notice that these pressure effects are almost completely reversible. On releasing the pressure down to zero, the PL signal indeed turns back toward the original one-peak shape although the position of the maximum appears at a wavelength lower than that originally measured at *P*=0, and the sample returns to be black at visual inspection. The general pressure behavior here observed is in a good agreement with that reported in ref. 10 apart from the slightly different position of the PL maximum after releasing the pressure down to zero. It is worth to notice that this weak extent of irreversibility shown by present PL measurements is well consistent with the above discussed structural behavior of the sample after releasing the pressure.

Looking at the frequency region around the excitation line, shown in the inset of Figure 9,

we can observe the emergence of a clear signal below 150 cm$^{-1}$ at 4.3 GPa, i.e. above the amorphization pressure threshold of about 3.3 GPa.

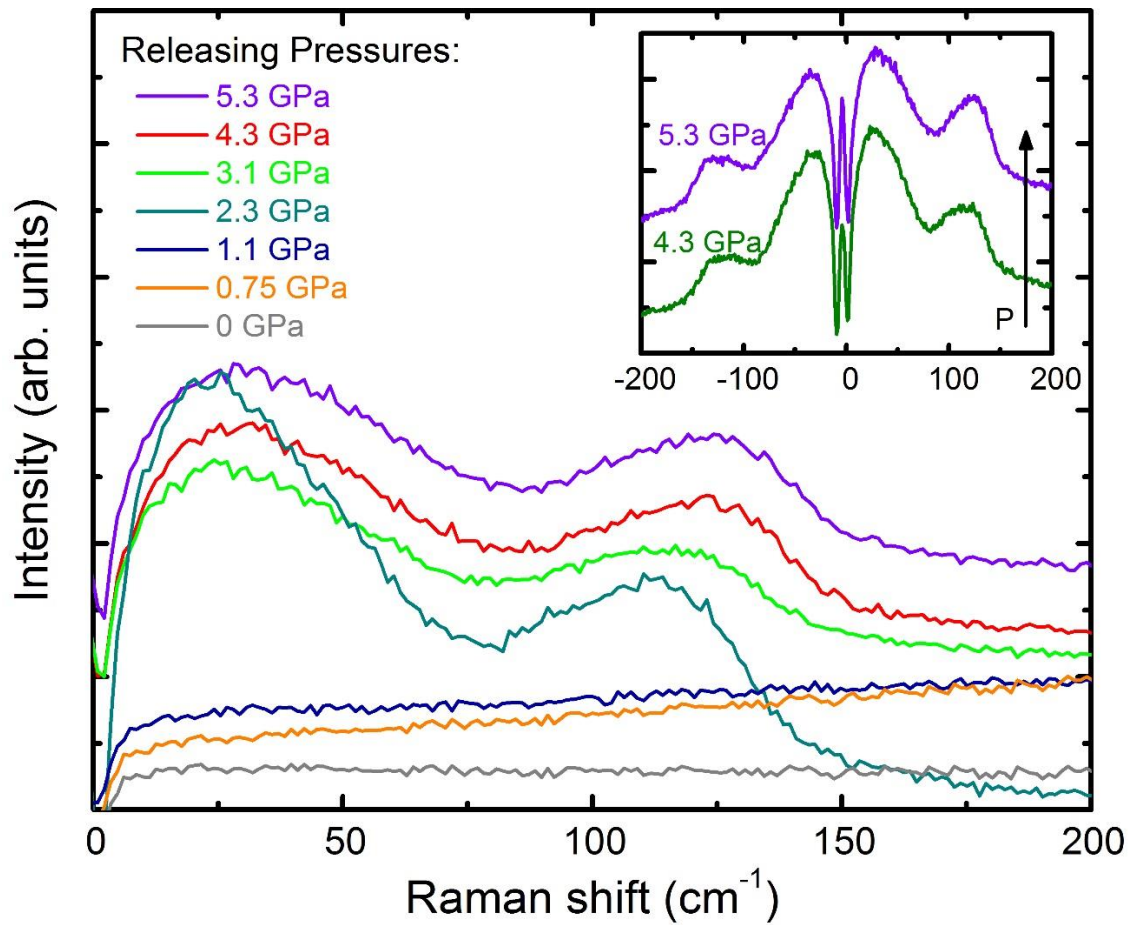

**Figure 9**: Low-frequency region of the Raman spectrum of MAPbI$_3$ on releasing pressure. The frequency region around the elastic line is reported in the inset.

The spectra in the inset are plotted in ($\nu_0 - \nu$), Raman shift, which clearly shows the intensity ratio expected for the Stokes and anti-Stokes Raman components. The spectral features, shown in greater detail in Figure 9, are therefore ascribed to a Raman signal. These appear only over the highest pressure range since the intensity of the fluorescence signal progressively reduces as the pressure is increased. The main Raman band, centered around 110 cm$^{-1}$ at 2.3 GPa, can be tentatively ascribed to Pb-I vibrations according to recent first-principles calculations.[19] The quite broad line shape, on the other hand, does not allow a more detailed vibrational assignment and indicates a rather high level of disorder.

**CONCLUSIONS**

In this paper we carefully analyzed the high-pressure behavior of the MAPbI$_3$ perovskite by means of HP XRD and PL. Structural data indicate the onset of a phase transition towards a orthorhombic structure at relatively low pressure (about 0.26 GPa) that remains stable up to the system amorphization which starts around 2.3 GPa and is completed (*i.e.*, full amorphous phase) around 3 GPa. The amorphous phase retains a partial stability by reducing pressure and the crystalline phase, found after pressure release, is the orthorhombic *Imm*2.

As to the PL spectra, it is clear that pressure deeply affects the electronic structure, resulting in a non trivial behavior of the emission bands of the system. While the new high pressure orthorhombic phase does not change remarkably the PL spectrum, the crystal amorphization (occurring above 3 GPa), first followed by the appearance of a new high energy peak, suggests the onset of a new mechanism related to light absorption. Support to this hypothesis is provided by the visual inspection of the micro-photographs, where the sample changes its color (from black to red) crossing the critical pressure. On further increasing the pressure, the suppression of the PL takes place, allowing for the observation of a weak Raman signal associated to Pb-I vibrations.

The present results open new experimental and theoretical demands on the role of electronic and structural degrees of freedom in driving the physical properties of this system that is representative of a large class of very appealing materials for device applications.

## SUPPLEMENTARY MATERIALS

See supplementary material for details about the lattice parameters derived from HP XRD analysis, thermal analysis and SEM analysis of MAPbI$_3$.

## AUTHOR INFORMATION

The authors declare no competing financial interests.

## ACKNOWLEDGMENT

We acknowledge ESRF for provision of beamtime.